\documentclass[aps,pre,groupedaddress,showpacs]{revtex4}
\usepackage{epsf}

\newcommand{\be}{\begin{equation}}
\newcommand{\ee}{\end{equation}}
\newcommand{\ba}{\begin{eqnarray}}
\newcommand{\ea}{\end{eqnarray}}
\newcommand{\ed}{\end{document}}
\newcommand{\lab}[1]{\label{#1}}
\newcommand{\re}[1]{(\ref{#1})}
\newcommand{\ci}[1]{\cite{#1}}
\newcommand{\bfr}{\begin{flushright}}
\newcommand{\efr}{\end{flushright}}
\newcommand{\bfl}{\begin{flushleft}}
\newcommand{\efl}{\end{flushleft}}
\renewcommand{\baselinestretch}{1.2}
\date{}
 
\begin{document}
 \title{CHAOTIZATION  OF THE PERIODICALLY DRIVEN QUARKONIA  }
\author{D.U.Matrasulov$^{(a)}$,
F.C.Khanna$^{(a)}$ and D.M.Otajanov$^{(b)}$}
\date{\today}
\affiliation{
(a) Physics Department University of Alberta\\
 Edmonton Alberta, T6G 2J1 Canada\\
 and TRIUMF, 4004 Wersbrook Mall,\\ Vancouver, British Columbia, Canada, V6T2A3\\
(b) Heat Physics Department of the Uzbek Academy of
Sciences,\\
28 Katartal St.,700135 Tashkent, Uzbekistan}

\begin{abstract}
Classical regular and chaotic dynamics of the particle bound in the Coulomb plus linear potential
under the influence of time-periodical perturbations is treated using resonace analysis.
Critical value of the external field at which chaotization will occur
is evaluated analytically based on the Chirikov criterion of stochasticity.
\end{abstract}
\maketitle

\section{Introduction}
Theory of dynamical chaos is one the rapidly developing fields
of the contemporary  physics.
It has found applications not only in various areas of physics but also in
many other areas natural sciences. Therefore the deterministic, or dynamical chaos is becoming the subject of
extensive theoretical as well as experimental investigation now \ci{chir}-\ci{del83}.

Dynamical chaos  is a phenomenon peculiar to the dynamical (deterministic) systems,
whose motion in some state space  is completely determined by a given interaction and the initial conditions.
However, under appropriate conditions, the behavior of these systems are indistguishible from random behavior despite
the absence of noise or thermal fluctuations.  Among the  theoretical models used to study and illustrate chaotic behavior
are various confined billiards geometries \ci{eckh88}, nonlinear oscillators\ci{chir,esca85,eckh88}, highly excited atoms
in mono- and poly-chromatic fields \ci{jens,koch,del83},
kicked rotator\ci{Izr90}, and a number
of astrophysical systems\ci{zas88}. In recent years a wide variety of experimental systems have emerged in which  simple theoretical paradigms
have been realized and which provide controlled environments to test and apply ideas.
One of the main features of dynamical chaos is the strong sensitivity of the evolution of the dynamical system with respect  to  changes
in the initial conditions. The small changes in the initial conditions may lead to considerable changes in the evolution.
As is well known \ci{chir,eckh88,Zas,jens84} chaotic motion occurs when the phase-space trajectories corresponding to neighboring resonances
overlap each other. This fact gives a criterion for the estimates of  stochasticity. One of the commonly used criterions of stochasticity,
Chirikov criterion, is based on this fact.

Dynamical systems which can exhibit chaotic dynamics can be divided into two classes:
time independent  and time-dependent systems. Billiards, atoms in a constant magnetic field, celestial systems with chaotic dynamics
are time independent systems, whose dynamics
can be chaotic.
Periodically driven rotators and  pendula, atoms and molecules in a microwave field,
 and many other periodicallyy driven dynamical systems are time-dependent
systems, whose dynamics can be chaotic.
A convenient testing ground for the theoretical and experimental study of chaos in the time-dependent dynamical systems
is the highly excited hydrogen atom in a monochromatic field \ci{cas87,jens,koch,del83}.
A theoretical analysis, of the behaviour of a classical
hydrogen atom interacting with monochromatic field,
based on the resonance overlap criterion \ci{del83}-\ci{mat1}, shows that for some critical value of the external field strength
$\epsilon_{cr}$, the electron enters into chaotic regime of motion, marked by unlimited diffusion
along orbits, leading to ionization.  Experimentally this phenomenon  was first observed  by Bayfield and Koch \ci{bay74}.
Theoretical explanation of this phenomenon was given later by severeal authors \ci{del,del83,koch}.
Such an ionization was called chaotic \ci{cas87,jens} or diffusive \ci{del83} ionization.
During the last three decades  chaotic ionization of nonrelativistic atom
was investigated by many authors theoretically \ci{jens84,del83,cas87,cas88,del} as well as experimentally \ci{jens,bay74,koch}.

In this paper we consider the QCD counterpart of this problem. Namely, we address the problem of regular and chaotic
motion in periodically driven quarkonium. Using resonanse analysis based on the Chirikov criterion of stochasticity we
estimate critical values of the external field strength at which quarkonium motion enters into chaotic regime.

Quarkonium in a monochromatic field can be considered as an analog of the hydrogen atom in a monochromatic field, in which
Coulomb potential is replaced by Coulomb plus confining potential.
Quarkonia   have been the subject of extensive experimental as well as theoretical studies for the last two decades \ci{AL}-\ci{Zhu}.
In the framework of potential approach the description of quark motion in  hadrons is reduced to solving
classical or quantum mechanical equations of motion  with Coulomb plus confining potential.
Considerable progress has been made in the calculation of the spectra of quarkonium \ci{Quig}-\ci{Zhu}.
At present a large amount of theoretical and experimental data on quarkonium properties is available.
Due to the presence of the confining potential quarkonium motion is equivalent to the atom in an uniform magnetic field whose dynamics
is also chaotic \ci{eckh88,Wint}.
 Recently chaotic dynamics  in hadrons and QCD has become a subject of theoretical studies \ci{Gu}-\ci{Buk}.
In particular, it is found that QCD is governed by quantum chaos in both confined and deconfined
phases \ci{Hal,Bit}.
The statistical analysis of the measured meson and baryon spectra shows that there is quantum chaos phenomenon
in these systems \ci{Pasc}.
The  study of the charmonium spectral statistics  and its dependence on color screening has established quantum chaotic behaviour
\ci{Gu}. It was claimed that such a behaviour could be the reason for  $J/\Psi$ supression \ci{Gu}.
Study of the chaotic dynamics of the  periodically driven quarkonium is needed  due to the recent advances in the creation of hot and dense matter,
such as hadronic matter and quark-gluon plasma.
Being in a quark gluon plasma or  in  hadronic matter quarkonium can be subjected to the influence of time-dependent fields, that
can lead to chaotization of the quarkonium and level fluctuations.
Indeed, the study of the spectral statistics of the charmonium based on the solution of the Bethe-Salpether
equation for the potential with color screening
shows that
regular motion can be expected at a small values of color screening mass but the chaotic motion is expected at a large one \ci{Gu}.
Periodically driven quarkonium can be also realized in the interaction of mesons with laser fields.
It should be noted, that  being a confined system quarkonium can exhibit  chaotic motion
even in the absence of any exterenal force.
It is well known that many realistic and model confined dynamical systems as an atom in a magnetic field\ci{Wint},
particle motion in resonators \ci{ding}, quantum dots\ci{alh} and
various billiards \ci{Zas}  can exhibit chaotic dynamics.
In next section we will treat a simple model, a one-dimensional quarkonium under a periodic perturbation.
In section 3 we extend our treatment to the three-dimensional case.
Some concluding remarks are given in section 4.

\section{One-dimensional model}

For  simplicity we consider a one-dimensional model with a potential
$$
V(x)=\left\{\begin{array}{ll}-\frac{Z}{x}+\lambda x\, & for\;\:x>0 \\ \\
\infty \, & for\:\;x \leq 0\end{array} \right.\,, $$

where $Z = \frac{4}{3}\alpha_{s}$ $\alpha_s$ being the strong coupling constant and $\lambda$ gives strength of the confining potential.
As is well known, in the case of the hydrogen atom interacting with a monochromatic field,
one-dimensional model provides an excellent description of the experimental chaotization
thresholds for real three-dimensional hydrogen atom \ci{jens,jens84,cas87}. The same success is to be expected
in the case of quarkonium.

The unperturbed Hamiltonian for the above potential is
\be
H_0 = \frac{p^2}{2} -\frac{Z}{x} +\lambda x .
\lab{unper}
\ee

We will treat the interaction of the system given by Hamiltonian \re{unper} with the periodic external potential of the form
\be
U(x,t) = \epsilon x cos\omega t.
\lab{pert}
\ee
with $\epsilon$ and $\omega$ being the field strength and frequency, respectively.
Thus the total Hamiltonian of the periodically driven quarkonium is
\be
H =H_0 + U(x,t)
\lab{total}
\ee

Formally the Hamiltonian \re{unper}  is equivalent to the Hamiltonian of the hydropen atom in constant homogenious
electric field. Chaotic dynamics of hydrogen atom in constant electric field under the influence of time-periodic field
was treated earlier \ci{ber,bala}.
To treat  nonlinear dynamics of this system under the influence of periodic perturbations
we  need to rewrite \re{unper} in action-angle variables.
Action can be found using its standard definition:
\be
n(E) =\frac{1}{2\pi} \int\limits_{c}^{a} pdx = \sqrt{2\lambda}\int\limits_{c}^{a}
\sqrt{\frac{(a-x)(x-c)}{x}}dx,
\ee
where
the momentum $p$ is given by
\be
p= \sqrt{2(E-V(x))},
\ee
and the constants $a$ and $c$  given as
$$
a = \frac{E+\sqrt{E^2+2Z\lambda}}{2\lambda},\;\;\;
c = \frac{E-\sqrt{E^2+2Z\lambda}}{2\lambda}
$$
are turning points of the particle. Since $c<0$ for the action we have
$$
n(E) = \frac{1}{2\pi}\int\limits_{0}^{a} pdx =
$$
\be
= B\sqrt{a+\frac{1}{a}}\cdot\Biggl[\biggl(a-\frac{1}{a}\biggr)E(k)+
\frac{1}{a} K(k)\Biggr].
\lab{act1}
\ee
where
$$
B= \frac{2\sqrt{2}}{3\pi\lambda^\frac{1}{4}},
$$
here $E(k)$ and $E(k)$ are the elliptic integrals\ci{abr} and

\be
k^2 = \frac{a^2}{a^2+1} - (1+\frac{1}{a^2})^{-1}.
\ee

Here we consider the two limit cases:
$a \gg 1$ and  $a \ll 1$.

For the first  case ($a\gg 1$) we have:
\be
E \equiv H_0 = Z^2An^{2/3} \cdot\Biggl[1 - \frac{\lambda ln4B^{-2/3}n^{2/3}}{A^2n^{4/3}} \Biggr],
\ee
with
$$
A=\biggr( \frac{3\pi \lambda}{2\sqrt{2}}^\frac{2}{3}\biggl).
$$

The corresponding proper frequency is
\be
\omega_0 = \frac{2}{3}Z^2 \biggr[
\frac{A}{n^{1/3}} + \frac{\lambda}{An^{5/3}}\bigr[\ln(4A\sqrt{\lambda}n^{2/3})\bigl]-1 \biggl]
\lab{freq}
\ee

For the case $a\ll 1$ we have:
\be
E \equiv H_0 = 0.5 Z^2 (9.7\lambda n^2 - n^{-2});
\ee

The proper frequency for this Hamiltonian is
\be
\omega_0 = Z^2 (n^{-3}+9.7n\lambda).
\ee

The total Hamiltonian can be written as
\be
H = H_0 + \epsilon \sum x_k cos(k\theta -\omega t),
\lab{full}
\ee

with

\be
x_k(n) = - \int_{0}^{2\pi}x(n,\theta) e^{ik\theta}d\theta
\ee
being Fourier amplitude  of the  perturbation.
For $a\ll 1$ we have an estimate for the Fourier component
\be
x_k(n)\approx-\frac{4E(n)}{\lambda}
\frac{1}{k}
\sin^2\frac{\pi k \sqrt{\lambda}}{2}.
\ee
For  $a\gg 1$ we have
\be
x_k(n)=-\frac{2An^{2/3}}{\pi^2\lambda k^2}.
\ee

It  is well known that  the phase-space trajectories of the regular motion lie on tori(so-called KAM tori).
According to Kolmogorov-Arnold-Moser theorem for sufficiently  small fields most of the trajectories remain regular.
If the value of the external perturbation exceeds  some value, which is called the critical field strength, KAM tori  start to break down
and chaotization of the motion will occur \ci{zas88}.
In Fig. 1  the phase-space portrait of the periodically driven quarkonium is plotted for the following  values of parameters:
$Z=0.15$, $\lambda=0.4$, $\omega =10^{-5}$ and $\epsilon =4\cdot 10^{-4}$. Eight  regular and two chaotic
trajectories are shown. The values of these parameters are written in the systems of units where $m_q=\bar h=1$, where $m_q$ is quark mass,
$c$ is the light speed. The values are chosen to have chaotic as well as regular behaviour.
Fig.2 represents the pahse-space portrait of the periodically driven quarkonium for $a \ll 1$ case for the values of parameters
$Z=0.15$, $\lambda=10$, $\omega =10^{-5}$ and $\epsilon =4\cdot 10^{-3}$. Again, the system of units $m_q=\bar h=1$ is used.

To estimate the critical value  of the external filed strength $\epsilon_{cr}$
we use Chirkov's resonance overlap criterion \ci{chir,zas88,del83,jens,koch},
which can be written as:
\be
s = \frac{\Delta \nu_k +\Delta \nu_{k+1}}{\omega_{0}(k+1)-\omega_{0}(k)},
\lab{chir}
\ee
with
$$
\Delta \nu_k = (\frac{\epsilon x_k}{\omega_0'})
$$
being the width of the $k$-th resonance \ci{chir,zas88}
and
$$
\omega_0' = d\omega_0/dn.
$$
From the resonance condition we have
$$
{\omega_{0}(k)-\omega_{0}(k+1)} = \frac{\omega}{k}-\frac{\omega}{k+1} =
\frac{\omega}{k(k+1)}.
$$
Applying this criterion to the quarkonium Hamiltonian \re{full} we have for $a \gg1$

$$
\epsilon_{cr} = \frac{ 0.07 Z^2 \omega \pi^2 \lambda}{ n^2}
\cdot \frac{k (k+1)}{(k+1)^2 + k^2} \times
$$
\be
\left\{
 1  + \frac{ \lambda }{ A^2 n^\frac{4}{3} }
\left[ 5 \ln \left(  4 A \lambda^{-\frac{1}{2}}
n^\frac{2}{3} \right) - 7 \right]
\right\}
\lab{larg}
\ee

and for  $a\ll 1$:

$$
\epsilon_{cr} = \frac{0.3 \omega \alpha}{k(k+1) n^2}\cdot \frac{29\alpha n^4-9}{29\alpha n^4-3}\times
$$
\be
\times \left[ \frac{1}{k} \sin^2 ( k \sqrt{\alpha} \frac{\pi}{2})+
\frac{1}{k+1}  \sin^2 ( (k+1) \sqrt{\alpha} \frac{\pi}{2})
\right]^{-1}
\lab{smal}
\ee

In Table 1 the values of the critical field for both approximations for
$u\bar u$, $d\bar d$, $s\bar s$, $c\bar c$ and $b\bar b$ quarkonia at the
following values of parameters: $\omega =10^{9}Hz$, $\alpha_s =0.112$\ci{PDG}, $\lambda =0.2Gev^2$, $n=10$.
are presented. For light ($u,d,s$) quarkonia we use formula \re{larg} and formula \re{smal}
for bottomium and charmonium.

\section{Three-dimensional model}

The Hamiltonian for the three-dimensional model is
$$
H_0 = \frac{p^2_r}{2} - \frac{Z}{r} +\lambda r +\frac{L^2}{r^2}.
$$
where $L$ is the orbital angular momentum and $p_r$ is the radial momentum.

The action can be expressed in terms of  elliptic integrals \ci{seet}:
$$
n = \int\limits_{c}^{a} p dr = \int\limits_{c}^{a}
\sqrt{2(E- \frac{L^2}{r^2}+ \frac{Z}{r} -\lambda r )} dr=
$$
$$\left[
(2Z/3 -L^2/c +Ec/3) K(k) +E(a-c)/3 E(k) + \right.
$$
\be
\left.+ L^2 (c^{-1}-b^{-1}) \prod(\beta^2 , k)
\right]g/\sqrt{\lambda}
\lab{act2}
\ee

with $a$ and $c$ being  the turning points,
$K,E,\prod$ are complete elliptic integrals of the first, second and third kind, respectively \ci{abr}, and the constants are given as
$$
k^2 = (a-b)/(a-c),
$$
$$
\beta = c k^2 / b,
$$
$$
g=2/\sqrt{a-c}.
$$
From eq.\re{act2}  the unperturbed Hamiltonian as a function of $n$ can be found approximately for $E/\lambda \gg1$,
(which corresponds to $n \gg 1$
or large quarkonium masses):
$$
H_0=(\frac{3}{2} \lambda n)^{2/3}\left[1+\frac{\pi L}{3 n} \right].
$$
The proper frequency is
$$
\omega_0 = \frac{\partial H_0}{\partial n} =(\frac{2\lambda^2}{3})^{1/3}.
\left[
n^{-1/3} -\frac{\pi L}{6}n^{-4/3}
\right]
$$

Then the Hamiltonian of the three-dimensional quarkonium in a monochroamtic field can be written as
$$
H=H_0 + \epsilon a \cos (\omega t) \times
$$
\be
\{ -\frac{3}{2} e \sin \varphi +
2 \sum [ x_k \sin\psi \cos k\phi + y_k \cos\psi \sin k\phi ] \},
\ee

where
$$
x_k = \frac{2i}{\omega_0 kT}\int e^{i\omega_0 kt}\dot x dt, \;\;\;\;y_k = \frac{2i}{\omega_0 kT}\int e^{i\omega_0 kt}\dot y dt,
$$
and $\psi$ and $\phi$ are the Euler angles.
Again, using the  resonance overlap criterion \re{chir} in which the resonance width is defined by
$$
\Delta \nu_k = (\frac{\epsilon r_k}{\omega_0'}),
$$
where
$$
r_k =\sqrt{x_k^2+y_k^2},
$$
we obtain an estimate for the critical field:

$$
\epsilon_{cr}= \frac{0.07 \alpha\omega }{ k(k+1)\pi n^2 }
\left( 1 - \frac{\pi L }{n} \right)
$$
$$
\left\{
\sqrt{  \frac{16 \pi^2}{9}+\frac{1}{k^2} }
+ \sqrt{ \frac{16 \pi^2}{9}+\frac{1}{(k+1)^2} }
\right\}^{-1}
$$
\be
\times
\left[ 1 - \frac{L^2 }{4 \pi^4 n^2}  \right].
\ee

This estimate for the critical field assumes that $n$ is large.
If the external field strength has  the value exceeding $\epsilon_{cr}$, breaking of KAM surfaces in the pase space will occur and
and the quarkonium diffuses in action and  the motion becomes chaotic.

\section{Conclusions}
Summarizing we have treated the chaotic dynamics of  the quarkonium in a time periodic field.
Using the Chirikov's resonance overlap criterion we obtain estimates for the critical value of the external field
strength at which chaotization of the quarkonium motion will occur. The experimental realization of the
quarkonium motion under time periodic perturbation could be performed  in several cases:
in laser driven mesons and  in quarkonia in the hadronic or quark-gluon matter.
The quarkonium, being the QCD analog of the hydrogen atom  can be also considered as a confined atom.
However, as is seen from the above treatment,  in contrast to the case of periodically driven atom, where the absolute value
of the energy decreases  diffusively, the energy of the quarkonium in a monochromatic field grows by diffusive law.
\vskip0.5cm

\section{Acknowledgements}
The work of DUM is supported  by NATO Science Fellowship of Natural
Science and Engineering Research Council of Canada (NSERCC).
The work of FCK is supported by NSERCC.
The work of DMO is supported by a Grant of the  Uzbek Academy of Sciences  (contract No 33-02).

\newpage
\begin{center}
Table 1. The values of the critical field strength for various quarkonia.
\vskip 0.5 cm

\begin{tabular}{|c|c|c|c|c|c|c|} \hline
\multicolumn{1}{|c|}{No}&
\multicolumn{1}{|c|}{Quarkonium} &
\multicolumn{1}{|c|}{ Quark mass (in MeV)} &
\multicolumn{3}{|c|}{ Critical field (V/cm)}\\
\cline{4-6}
&&&$n=5$ & $n=7$  & $n=10$ \\
\hline
1 & $u\bar u$ & $5 $ & $2.124\cdot 10^8$& $1.084\cdot 10^8$ & $5.31\cdot 10^7$\\
\hline
2 & $d\bar d$ & $10$ & $ 2.231\cdot 10^{5}$& $1.138\cdot 10^{5}$ & $5.578\cdot 10^{4}$ \\
\hline
3 & $s\bar s$ & $150$ & $1.047\cdot 10^5$& $5.34\cdot 10^4$ & $2.617\cdot 10^4$ \\
\hline
4 & $c\bar c$ & $1500$ & $ 49.167$& $25.045$ & $12.258$ \\
\hline
5 & $b\bar b$ & $4800$ & $1.224$& $0.624$ & $0.306$ \\
\hline
\end{tabular}
\end{center}
\newpage

\begin{figure}[htb]
\begin{center}
   \epsfysize=10cm
\epsffile{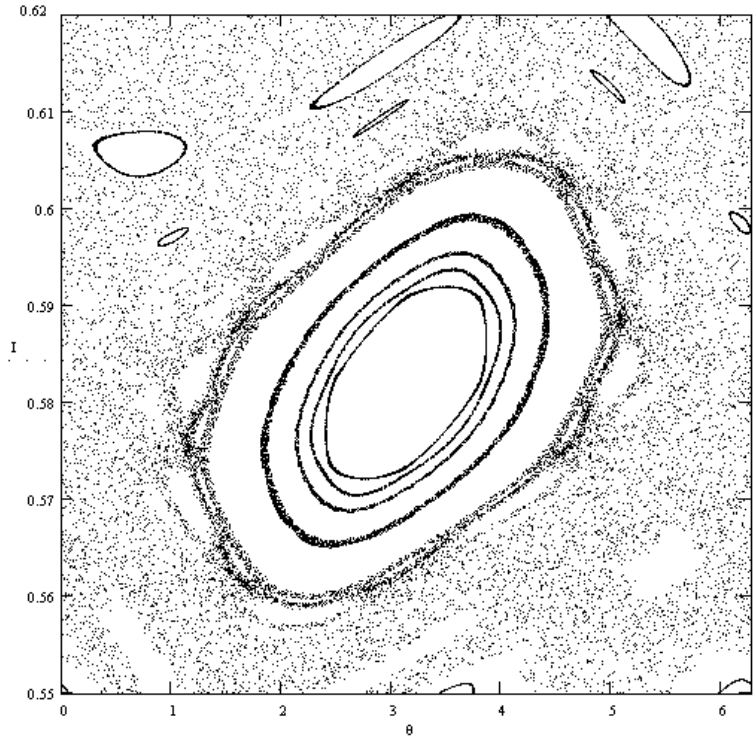}
\end{center}
   \vspace{.5cm}
\caption{
\label{3}
 Phase space portrait of the periodically driven quarkonium: $a>>1$ case.}
\end{figure}

\newpage
\begin{figure}[htb]
\begin{center}
   \epsfysize=10cm
\epsffile{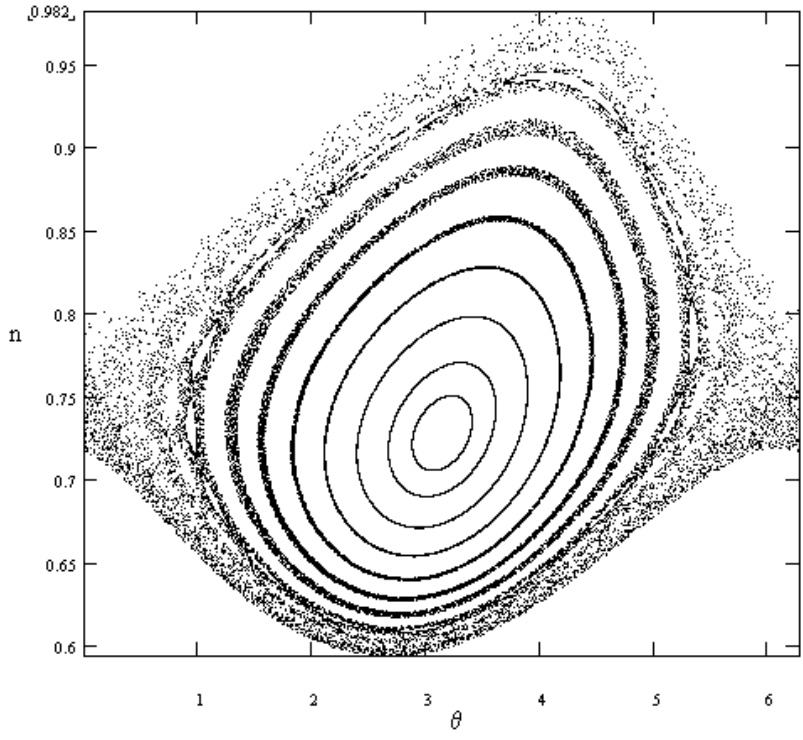}

\end{center}
   \vspace{.5cm}
\caption{
\label{3}
 Phase space portrait of the unperturbed quarkonium: $a<<1$ case.}
\end{figure}

\ed